\documentclass[aps,pre,
superscriptaddress,groupedaddress,showpacs,twocolumn]{revtex4}
\usepackage{graphicx,epsf}
\usepackage{xcolor}
\usepackage{psfrag}
\usepackage{psfrag}
\begin{document}
\title{Probability of loops formation in star polymers in long range correlated disorder}
\author{K. Haydukivska}
\affiliation{Institute for Condensed
Matter Physics of the National Academy of Sciences of Ukraine,\\
79011 Lviv, Ukraine}
\author{V. Blavatska}
\email[]{E-mail:  viktoria@icmp.lviv.ua}
\affiliation{Institute for Condensed
Matter Physics of the National Academy of Sciences of Ukraine,\\
79011 Lviv, Ukraine}

\begin{abstract}
We analyze the statistics of loops formation in $f$-branched star polymers in an environment with structural defects,
correlated at large distances $r$ according to a power law $\sim  r^{-a}$. Applying the direct polymer renormalization approach,
we found the values of the set of universal exponents, governing the scaling of probabilities of various types of loops in macromolecules.
\end{abstract}
\date{\today}

\maketitle

\section{Introduction}

Star-like polymers are attracting a lot of attention over last decades due to their important role in biophysics, medicine and technologies \cite{Hadjichristidis12}. These objects are used e.g. in
 drug delivery \cite{Zhu06},  bio-engineering \cite{Liu11},  and can be encountered in complex systems such
as gels, rubber or micellar systems \cite{Grest96,Likos01}.  Note that the star polymers can be considered as building blocks of polymer networks of a more complex structure \cite{Duplantier89}.

\begin{figure}[b!]
\begin{center}
\includegraphics[width=40mm]{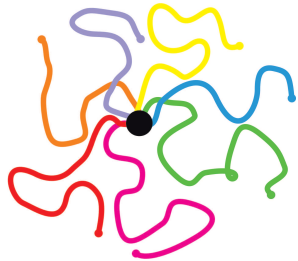}
\caption{ \label{fig:1} Schematic presentation of star polymer with $f{=}7$ arms. }
\end{center}\end{figure}

The star polymer can be considered as a set of linear chains (branches) that are connected by one of the ends to the core (see Fig.  \ref{fig:1}). In general, the star polymers are characterized by more compact structure and lover viscosity in solutions as comparing with linear chains of the same molecular weight \cite{Alexandros15,Hadjichristidis12,vonFerber}, which among other unique properties make them the interesting objects to study.

\begin{figure}[t!]
\begin{center}
\includegraphics[width=70mm]{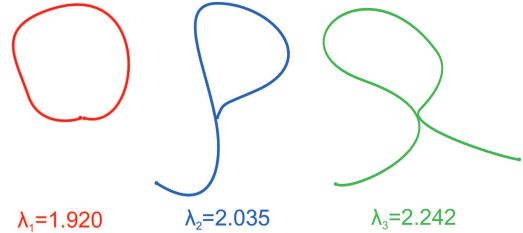}
\caption{ \label{fig:2} Schematic presentation of loops with different positions along the polymer chain and corresponding numerical
values of scaling exponents \cite{Hsu04}. }
\end{center}\end{figure}

It is well known that the loop formation in macromolecules plays an important
role in a number of biochemical processes, such as stabilization of globular proteins  \cite{Perry84,Wells86,Pace88,Nagi97},
transcriptional regularization of genes \cite{Schlief88,Rippe95,Towles09},
DNA compactification in the nucleus \cite{Fraser06,Simonis06,Dorier09} etc. The loop
corresponds to a “contact” between two monomers $i$
and $j$, which can be separated by a large distance along the polymer chain.
The probability to find a loop of a size $|i-j|$  obeys a scaling law\cite{Jacobson50}:
\begin{equation}
P \sim |i-j|^{-\lambda}, \label{probloop}
\end{equation}
where $\lambda$ is a universal exponent that depends only on global characteristics like space dimension $d$. In the idealized case of a Gaussian chain one has $\lambda=d/2$. However, {  when one takes into account} the excluded volume interactions between monomers,
the probability of loop formation strongly depends on position of monomers $i$ and $j$ along the chain \cite{Chan89,Redner80,Hsu04}.
In particular, it was found that the end monomers of a chain have a higher
probability of contact then the inner ones (see Fig. \ref{fig:2}). Note, that an interesting question of how the complex (branched) structure of polymer macromolecule influences the probability of loop formation is still not solved and motivated our present study.

It is important to notice that in the real physical situations one often encounters polymers in solutions in presence of another objects (impurities), like colloidal particles \cite{Pusey86} or various biochemical species in the crowded environment of biological cells \cite{Kumarrev,Minton01}. In the simplest case one can treat these objects as uncorrelated point-like defects. Then, for the small concentrations of obstacles it was proven both analytically \cite{Kim83} and numerically \cite{Kremer,Lee88,Woo91}, that conformational properties of polymer macromolecules do not change in comparison with the pure environment, unless the concentration of such impurities reaches the percolation threshold   \cite{Kremer,Grassberger93,Ordemann02,Janssen07}.
Completely different situation occurs when obstacles cannot be treated as point-like objects but form the complex fractal structures (clusters) \cite{Dullen79}.  Such situation  can be described within the frames of a model of long-range correlated disorder, where the defects are assumed to be correlated at large distances $r$ according to a power law resulting in a pair correlation function \cite{Weinrib83}:
\begin{equation}
g(r)\sim r^{-a}.\label{corfun}
\end{equation}
The correlation parameter $a<d$ is another global characteristic of the system. The impact of long-range correlated disorder on the conformational properties of   linear star-like and circular polymers has been studied analytically in the works  \cite{Blavatska01,Blavatska06,Haydukivska14}. In our previous study \cite{Haydukivska16} we analyzed the probability of a single loop formation in a linear polymer chain in environment with long-range correlated disorder
and obtain the values of critical exponents $\lambda$ governing the scaling of probabilities (\ref{probloop}). Our present paper thus aims to extend this study to the case of loop formation in $f$-branched star polymers in long-range correlated disorder { , in particular a case of a single loop formation via interaction of two monomers on a star-like polymer}.

\begin{figure}[t!]
\begin{center}
\includegraphics[width=85mm]{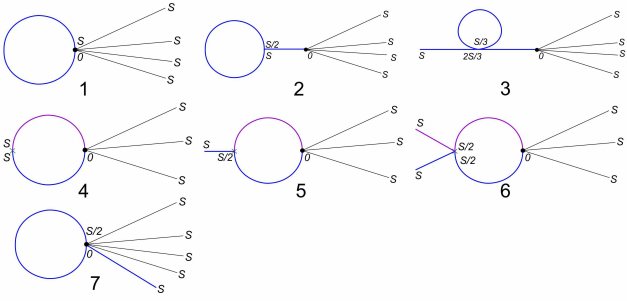}
\caption{ \label{fig:3} Schematic presentation of different possibilities of single loop formation  in a star polymer. }
\end{center}\end{figure}

The layout of the paper is as follows. In the next Section II we introduce the continuous model of $f$-branched star polymer.  Then in Section III we briefly introduce the method of direct polymer renormalization which is followed by  results for scaling exponents, obtained in Section IV. We end up with conclusions in Section V.

\section{The Model}

We study the statistics of loop formation in star polymers within the frames of continuous chain model \cite{Edwards}. We consider the star with $f$ branches each of equal length $S$. Each branch is presented by a  path  $\vec{r}_i(s)$,  $i=1,\ldots,f$, parameterized by  $0\leq s \leq S$.

There are {  7 } possibilities of a single loop formation in a star polymer (see Fig.  \ref{fig:3}). The partition function of the system  in each case reads:
\begin{eqnarray}
{\cal Z}_{y;f}^x(S)=\frac{ {\displaystyle{\int}}\! {\cal {D}}\vec{r}\,\,
\delta_y^x {\displaystyle {\prod\limits_{a=1}^{f}}}\delta(\vec{r}_a(0))
{\rm e}^{-H}}
{{\displaystyle {\int}}\! {\cal {D}}\vec{r}\,\,{\displaystyle{\prod\limits_{a=1}^{f}}}\delta(\vec{r}_a(0))
{\rm e}^{-H}}. \label{model-con}
\end{eqnarray}

Here, a multiple path integration is performed over the paths $\vec{r}_1,\ldots,r_f$, the product of $\delta$-functions $\delta(\vec{r}_i(0))$  reflects the star-like configuration of the system  and  $\delta_y^x$ {  ($x=1,2$, $y=1,2,3,4$) } describes the presence of a loop of a certain type, which is formed either by two monomers which belong to the same branch of macromolecule ($x=1$) or to the different branches $x=2$ (see Fig.  \ref{fig:3}):

\begin{eqnarray}
&&\delta_1^1=\sum_{a=1}^f\delta(\vec{r}_a(S)-\vec{r}_a(0)),\label{delta1}\\
&&{ \delta_2^1=\sum_{a=1}^f\delta(\vec{r}_a(S/2)-\vec{r}_a(0))},\\
&&\delta_3^1=\sum_{a=1}^f\delta(\vec{r}_a(S)-\vec{r}_a(S/2)),\\
&&\delta_4^1=\sum_{a=1}^f\delta(\vec{r}_a(2S/3)-\vec{r}_a(S/3)),\\
&&\delta_1^2=\sum_{a,b=1}^f\delta(\vec{r}_a(S)-\vec{r}_b(S)),\\
&&\delta_2^2=\sum_{a,b=1}^f\delta(\vec{r}_a(S)-\vec{r}_b(S/2)),\\
&&\delta_3^2=\sum_{a,b=1}^f\delta(\vec{r}_a(S/2)-\vec{r}_b(S/2)),
\label{delta6}
\end{eqnarray}
and $H$ is an effective Hamiltonian of the model:
\begin{eqnarray}
&&H=\sum_{a=1}^{f}\frac{1}{2}\int^{S}_{0}{\rm d}s\,\left(\frac{{\rm d}\vec{r}_a(s)}{ds}\right)^{2} +\nonumber\\
&&+\frac{u_0}{2}\sum_{a,b=1}^{f}\int^{S}_{0}{\rm d}s'\int^{S}_{0}{\rm d}s''\delta(\vec{r}_a(s')-\vec{r}_b(s''))-\nonumber\\
&&-\sum_{a=1}^{f}\int^{S}_{0}{\rm d}s\,V(\vec{r}_a(s)). \label{H-con}
\end{eqnarray}
The first term  in (\ref{H-con}) represents the chain connectivity,  the second describes the excluded volume interaction with a bare coupling constant $u_0$ and the last  term contains a random  potential $V(\vec{r}_a(s))$, which in this paper is considered to be characterized  by \cite{Weinrib83}:
\begin{eqnarray}
&&{\overline{ V(\vec{r}_a(s))}} =0,\nonumber\\
&& {\overline{ V(\vec{r}_a(s'))V(\vec{r}_b(s''))}} = w_0 |\vec{r}_a(s')-\vec{r}_b(s'')|^{-a}, \label{avv0}
\end{eqnarray}
where ${\overline{(\ldots)}}$ stands for the average over different realizations of disorder and $w_0$ is a corresponding coupling constant. Averaging partition functions (\ref{model-con}) over different realizations of disorder, taking into account up to the second moment of cumulant expansion and recalling (\ref{avv0}) we obtain the set of $ {\overline {{\cal Z}_{y;f}^x(S)}}$ with an effective Hamiltonian:
\begin{eqnarray}
&&H_{eff}=\sum_{a=1}^{f}\frac{1}{2}\int^{S}_{0}{\rm d}s\,\left(\frac{{\rm d}\vec{r}_a(s)}{ds}\right)^{2} +\nonumber\\
&&+\frac{u_0}{2}\sum_{a,b=1}^{f}\int^{S}_{0}{\rm d}s'\int^{S}_{0}{\rm d}s''\delta(\vec{r}_a(s')-\vec{r}_b(s''))-\nonumber\\
 &&-\frac{w_0}{2}\sum_{a,b=1}^{f}\int_0^{S}{\rm d}s'\int_0^{S}{\rm d}s{''}\, |\vec{r}_a(s'')-\vec{r}_b(s')|^{-a}.\nonumber
\label{Hdis}
\end{eqnarray}

To calculate the probabilities of loop formations (\ref{probloop}) one needs to consider the corresponding ratios for each type of loops presented on Fig.  \ref{fig:3}:
\begin{eqnarray}
&&P_{y}^{x}=\frac{{\overline {{\cal Z}_{y;f}^x(S)}}}{{\overline {{\cal Z}_{f}(S)}}}\sim S^{\lambda_y^x},   \label{P}
\end{eqnarray}
 where ${{\overline {{\cal Z}_f(S)}}}$ is the total partition function of a star polymer  given by:
\begin{eqnarray}
{\overline{ {\cal Z}_f(S)}}=\frac{ {\displaystyle{\int}}\! {\cal {D}}\vec{r}\,\,
{\displaystyle {\prod\limits_{a=1}^{f}}}\delta(\vec{r}_a(0))
{\rm e}^{-H_{eff}}}
{{\displaystyle {\int}}\! {\cal {D}}\vec{r}\,\,{\displaystyle{\prod\limits_{a=1}^{f}}}\delta(\vec{r}_a(0))
{\rm e}^{-H_0}} \label{star}
\end{eqnarray}
with $H_0=\sum_{a=1}^f\frac{1}{2}\int^{S}_{0}{\rm d}s\,\left(\frac{{\rm d}\vec{r}_a(s)}{ds}\right)^{2}$ is Hamiltonian of a Gaussian star polymer.

\section{The Method}

We aim to calculate the universal scaling exponents $\lambda_y^x$  governing the asymptotic behavior of probabilities of loop formation in star polymers.
For this purpose we apply a direct renormalization scheme developed by des Cloiseaux \cite{desCloiseaux}.
The method is based on the eliminations of various divergences, observed in the asymptotic limit of an infinitely long chains
(corresponding to an infinite of number polymer configurations) by absorbing them into the set of the so-called  renormalization factors.
  In particular, the renormalization factor $\chi_1(\{ z_0 \})$ (with $\{z_0\}$ being the set of bare coupling constants) is introduced via:
 \begin{eqnarray}
\frac{{\cal Z}_f(S)} { {\cal Z}_f^{{\rm Gauss}}(S)}=\chi_1(\{ z_0 \}).
   \end{eqnarray}
Here, ${\cal Z}_f(S)$ is the partition function of an $f$-branched star polymer  and
${\cal Z}^{{\rm Gauss}}_f(S)$ is the partition function of an idealized Gaussian model.
In a similar way, in our problem we may introduce the factors $\chi_y^x(\{z_0 \})$ via:
\begin{equation}
\frac{P_y^{x}}{{P_y^x}^{{\rm Gauss}}}=\chi_y^{x}(\{z_0 \}).
\end{equation}
Recalling the scaling of loop probabilities
(\ref{P}) and remembering the fact that
in Gaussian approximation ${P_y^x}^{{\rm Gauss}}\sim S^{-d/2}$,
we find the estimates for the effective critical exponents $\lambda_y^{x}(\{z_0 \})$:
\begin{equation}
\lambda_y^{x}(\{z_0 \})-d/2=-S \frac{\partial \ln \chi_y^{x}(\{z_0 \}) }{\partial S}.\label{nuexp}
\end{equation}

The critical exponents  presented in the form of series expansions in the coupling constants
are divergent in the asymptotic limit of large $S$.
 To eliminate these divergences, the renormalization of the coupling constants is performed.
The critical exponents attain finite values when evaluated at a stable fixed point  of the renormalization group transformation.
 The flows of the renormalized coupling constants $\{z_{\mathrm{R}}\}$ are governed by functions $\beta_{\{z_{\mathrm{R}}\}}$:
\begin{equation}
\beta_{\{z_{\mathrm{R}}\}}=2S\frac{\partial z_{\mathrm{R}}(\{ z_0\})}{\partial S}.
\end{equation}
 Reexpressing $\{ z_0\}$ in terms of renormalized couplings $\{z_{\mathrm{R}}\}$,  the fixed points of renormalization group transformations are given
 by common zeros of the $\beta$-functions.
 Stable fixed points govern the asymptotical scaling properties of macromolecules and gives
 reliable asymptotical values of the critical exponents  (\ref{nuexp}).

\section{Results}

In the simplified case of a Gaussian chain the expressions for the partition functions (\ref{model-con}) read:
\begin{eqnarray}
&&{\cal Z}_{y;f}^1(S)=\frac{ {\displaystyle{\int}}\! {\cal {D}}\vec{r}\,\,
\sum\limits^f_{a=1}\delta(\vec{r}_a(m)-\vec{r}_a(n)) {\displaystyle {\prod\limits_{a=1}^{f}}}\delta(\vec{r}_a(0))
{\rm e}^{-H_0}}
{{\displaystyle {\int}}\! {\cal {D}}\vec{r}\,\,{\displaystyle{\prod\limits_{a=1}^{f}}}\delta(\vec{r}_a(0))
{\rm e}^{-H_0}},\nonumber\\
&&{\cal Z}_{y;f}^2(S)=\frac{ {\displaystyle{\int}}\! {\cal {D}}\vec{r}
\sum\limits^f_{a,b=1}\delta(\vec{r}_a(m)-\vec{r}_b(n)) {\displaystyle {\prod\limits_{a=1}^{f}}}\delta(\vec{r}_a(0))
{\rm e}^{-H_0}}
{{\displaystyle {\int}}\! {\cal {D}}\vec{r}\,\,{\displaystyle{\prod\limits_{a=1}^{f}}}\delta(\vec{r}_a(0))
{\rm e}^{-H_0}},\nonumber
\end{eqnarray}
where $m$ and $n$ are the corresponding arguments of the functions (\ref{delta1}) - (\ref{delta6}), taking on the values $0$, $S/2$, $S/3$, $S$, $2S/3$.

Using Fourier transformation for the delta-functions:
\begin{eqnarray}
&&\delta(\vec{r}_a(m)-\vec{r}_b(n)) =\frac{1}{(2\pi)^{d}}\int {\rm d}\vec{q}\, {\rm e}^{-i\vec{q}(\vec{r}_a(m)-\vec{r}_b(n))}\nonumber
\end{eqnarray}
we receive
\begin{eqnarray}
&&{\cal Z}_{y;f}^1(S)=f(2\pi(m-n))^{-d/2}, \\
&&{\cal Z}_{y;f}^2(S)=\frac{f(f-1)}{2}(2\pi(m+n))^{-d/2},
\end{eqnarray}
where $m-n$ and $m+n$ corresponds to the loop length that can be $S/3$, $S/2$, $S$, $3S/2$ and $2S$ depending on the type of the loop under consideration.

\begin{figure}[b!]
\begin{center}
\includegraphics[width=80mm]{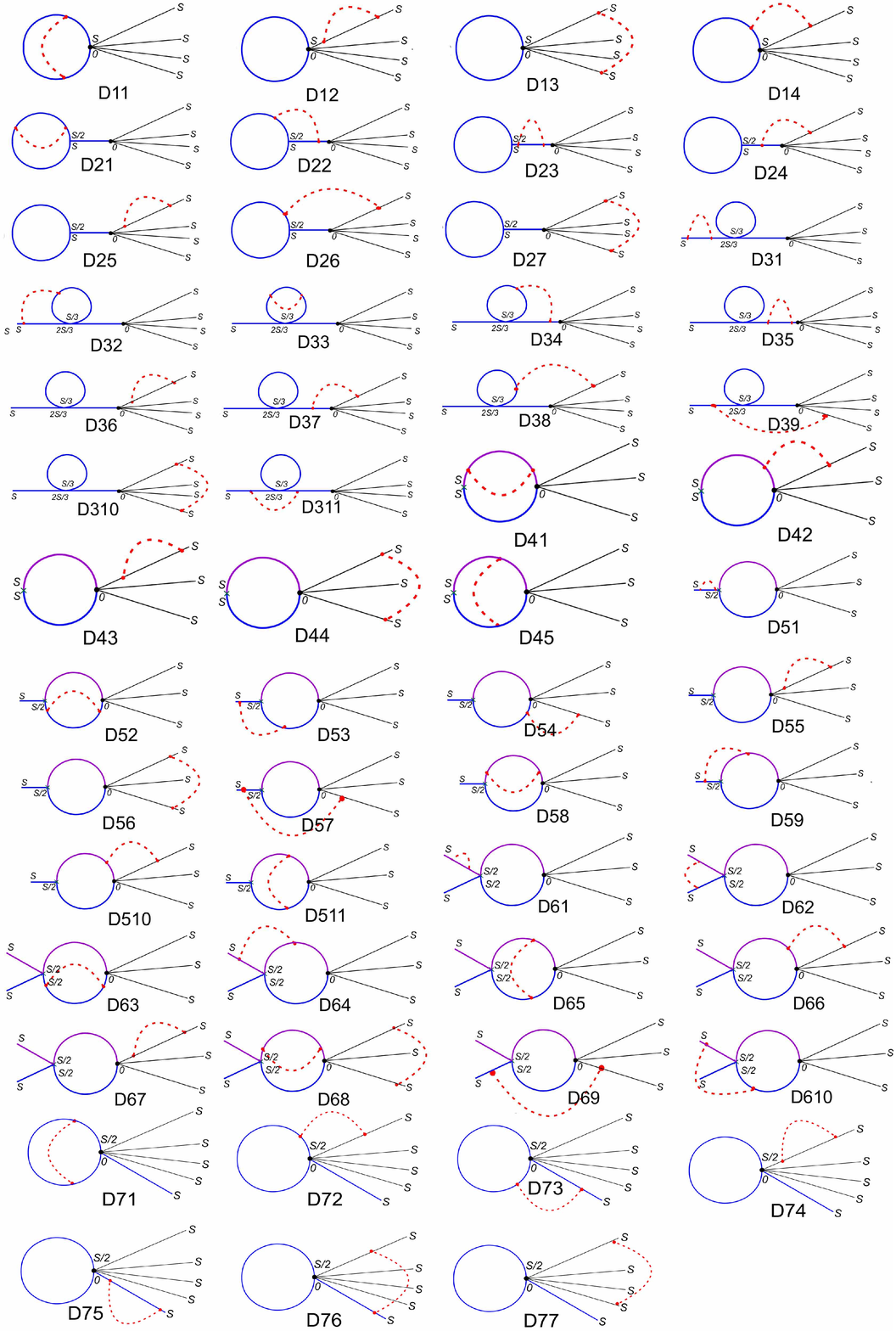}
\caption{ \label{fig:4} Diagrammatic presentation of contributions into ${Z_{y;f}^x}^u$ in the one loop approximation. }
\end{center}\end{figure}

 We consider the coupling constants $u_0$ and $w_0$ to be small so that we could present the  partition functions as the  perturbation theory series in these constants.
Keeping contributions up to the first order, in general the partition function can be presented as:
\begin{equation}
{\cal Z}_{y;f}^x(S)=\xi(2 \pi S)^{-d/2}-z_u{Z_{y;f}^x}^u+z_w{Z_{y;f}^x}^w+\ldots
\end{equation}
where $\xi$ is $f$ or $\frac{f(f-1)}{2}$  and $z_u{=}u_0(2\pi)^{-d/2}S^{2-d/2}$, $z_w{=}w_0(2\pi)^{-a/2}S^{2-a/2}$ are dimensionless coupling constants. To calculate ${Z_{y;f}^x}^u$ and ${Z_{y;f}^x}^w$, it is useful to apply the diagram technique and present the contributions as a sum of all possible diagrams (see Fig. \ref{fig:4}). { Note that diagrams D12, D14, D24, D25, D26, D36, D37, D38, D39,D72,D74,D76 are taken with combinatorial factor $(f-1)$; diagrams D43, D54, D55, D57, D510, D67 with $(f-2)$; D66, D69 and D42 with $2(f-2)$; D13, D27, D310,D77 with $(f-1)(f-2)/2$; D68, D56, D44 with $(f-2)(f-3)/2$ and diagrams D41, D61, D63, D64, and D610 are taken twice.} An example of diagram calculation can be found in Appendix A and the analytical expressions corresponding to
all the diagrams are presented in Appendix B. Here it is important to notice that corresponding expressions for  ${Z_{y;f}^x}^w$ can be obtained by simple  substitute of $d$ by $a$ in  ${Z_{y;f}^x}^u$ \cite{Haydukivska14}.

Performing a double expansion over $\varepsilon=4-d$ and $\delta=4-a$ and keeping only those terms that contain poles the partition functions can be presented as:
\begin{eqnarray}
&&{\cal Z}_{1;f}^1(S)= (2\pi)^{-d/2} S^{-d/2} f\left(1-z_u\frac{f^2-f+4}{\varepsilon}+\right.\nonumber\\
&&\left.+z_w\frac{f^2-f+4}{\delta}\right),\\
&&{ {\cal Z}_{2;f}^1(S)= (2\pi)^{-d/2} S^{-d/2} f\left(1-z_u\frac{f^2+f+4}{\varepsilon}+\right.}\nonumber\\
&&{ \left.+z_w\frac{f^2+f+4}{\delta}\right)},\\
&&{\cal Z}_{3;f}^1(S)= (2\pi)^{-d/2} \left(\frac{S}{2}\right)^{-d/2} f\left(1-\right.\nonumber\\
&&\left.-z_u\frac{f^2-3f+8}{\varepsilon}+z_w\frac{f^2-3f+8}{\delta}\right),\\
&&{\cal Z}_{4;f}^1(S)= (2\pi)^{-d/2} \left(\frac{S}{3}\right)^{-d/2} f\left(1-\right.\nonumber\\
&&\left.-z_u\frac{f^2-3f+12}{\varepsilon}+
z_w\frac{f^2-3f+12}{\delta}\right),\\
&&{\cal Z}_{1;f}^2(S)= (2\pi)^{-d/2} (2S)^{-d/2} \frac{f(f-1)}{2}\left(1-\right.\nonumber\\
&&\left.-z_u\frac{f^2-3f+6}{\varepsilon}+z_w\frac{f^2-3f+6}{\delta}\right),\\
&&{\cal Z}_{2;f}^2(S)= (2\pi)^{-d/2} \left(\frac{3S}{2}\right)^{-d/2}\frac{f(f-1)}{2} \left(1-\right.\nonumber\\
&&\left.-z_u\frac{f^2-3f+8}{\varepsilon}+z_w\frac{f^2-3f+8}{\delta}\right),\\
&&{\cal Z}_{3;f}^2(S)= (2\pi)^{-d/2} S^{-d/2} \frac{f(f-1)}{2}\left(1-\right.\nonumber\\
&&\left.-z_u\frac{f^2-3f+12}{\varepsilon}+z_w\frac{f^2-3f+12}{\delta}\right),\\
&&{\cal Z}_{f}(S)= 1-z_u\frac{f^2-3f}{\varepsilon}+z_w\frac{f^2-3f}{\delta}.
\end{eqnarray}

Using the definition of probability of loop formation (\ref{P}),  taking into account that $z_u$ and $z_w$ in the first order of perturbation theory
are proportional to $\varepsilon$, $\delta$, and keeping only contributions up to $\varepsilon$ and  $\delta$,  we receive the corresponding relations:
\begin{eqnarray}
&&P_1^1(S)= (2\pi)^{-d/2} S^{-d/2} f\left(1-z_u\frac{2f+4}{\varepsilon}+\right.\nonumber\\
&&+\left.z_w\frac{2f+4}{\delta}\right), \label{p1}\\
&&{ P_2^1(S)= (2\pi)^{-d/2} S^{-d/2} f\left(1-z_u\frac{4f+4}{\varepsilon}+\right.}\nonumber\\
&&{ +\left.z_w\frac{4f+4}{\delta}\right), }\label{p1}\\
&&P_3^1(S)= (2\pi)^{-d/2} \left(\frac{S}{2}\right)^{-d/2} f\left(1-z_u\frac{8}{\varepsilon}+z_w\frac{8}{\delta}\right),\nonumber\\
&&P_4^1(S)= (2\pi)^{-d/2} \left(\frac{S}{3} \right)^{-d/2} f\left(1-z_u\frac{12}{\varepsilon}+z_w\frac{12}{\delta}\right),\nonumber\label{p3}\\
&&P_1^2(S)= (2\pi)^{-d/2} (2S)^{-d/2} \frac{f(f-1)}{2}\left(1-\right.\nonumber\\
&&\left.-z_u\frac{6}{\varepsilon}+z_w\frac{6}{\delta}\right),\nonumber\\
&&P_2^2(S)= (2\pi)^{-d/2} \left(\frac{3S}{2}\right)^{-d/2}\frac{f(f-1)}{2} \left(1-z_u\frac{8}{\varepsilon}+\right.\nonumber\\
&&\left.+z_w\frac{8}{\delta}\right),\nonumber\\
&&P_3^2(S)= (2\pi)^{-d/2} S^{-d/2} \frac{f(f-1)}{2}  \left(1-z_u\frac{12}{\varepsilon}+z_w\frac{12}{\delta}\right).\nonumber
\end{eqnarray}

Recalling (\ref{nuexp}) we find the analytical expressions for the scaling exponents $\lambda_y^x$ in the form of series in couplings $z_u$, $z_w$:
\begin{eqnarray}
&&\lambda_{1}^1-d/2= (f+2)z_u-(f+2)z_w,\label{l1}\\
&&{ \lambda_{2}^1-d/2= (2f+2)z_u-(2f+2)z_w},\\
&&\lambda_{3}^1-d/2= 4z_u-4z_w,\\
&&\lambda_{4}^1-d/2= 6z_u-6z_w,\\
&&\lambda_1^2-d/2= 3z_u-3z_w,\\
&&\lambda_2^2-d/2= 4z_u-4z_w,\\
&&\lambda_3^2-d/2= 6z_u-6z_w.\label{l6}
\end{eqnarray}
The further analysis of these expressions is based on the values of fixed points of a polymer in long-range correlated disorder, found previously in Ref. \cite{Blavatska01}:
\begin{eqnarray}
&& {\rm {Gaussian}}: z^*_{u_0}=0,  z^*_{w_0}=0, \label{FPG}\\
&& { \rm {Pure}}: z^*_{u_0}=\frac{\varepsilon}{8}, z^*_{w_0}=0, \label{FPP} \\
&& {\rm  {LR}}: z^*_{u_0}=\frac{\delta^2}{4(\varepsilon-\delta)},  z^*_{w_0}=\frac{\delta(\varepsilon-2\delta)}{4(\delta-\varepsilon)}. \label{FPL}
\end{eqnarray}
Here,  (\ref{FPG}) corresponds to an idealized Gaussian macromolecule in pure environment, (\ref{FPP})
describes polymer with excluded volume interaction, and finally (\ref{FPL}) corresponds to the case of polymer in solution  with obstacles correlated on large distances according to (\ref{corfun}). Evaluating expressions  (\ref{l1})-(\ref{l6}) in Gaussian fixed point, we simply restore the known result $\lambda_y^x=d/2$. For the polymers with excluded volume interaction in pure solvent we obtain:
\begin{eqnarray}
&&{\lambda_1^1}^{{ \rm {Pure}}}=2+\frac{f-2}{8}\varepsilon,\label{exp1}\\
&&{ {\lambda_2^1}^{{ \rm {Pure}}}=2+\frac{f-1}{4}\varepsilon,}\label{exp11}\\
&&{\lambda_3^1}^{{ \rm {Pure}}}=2,\\
&&{\lambda_4^1}^{{ \rm {Pure}}}=2+\frac{\varepsilon}{4},\\
&&{\lambda_1^2}^{{ \rm {Pure}}}=2-\frac{\varepsilon}{8},\\
&&{\lambda_2^2}^{{ \rm {Pure}}}=2,\\
&&{\lambda_3^2}^{{ \rm {Pure}}}=2+\frac{\varepsilon}{4}.
\end{eqnarray}
{  Note, that only the exponents ${\lambda_1^1}^{{ \rm {Pure}}}$ and ${\lambda_2^1}^{{ \rm {Pure}}}$ are nontrivial and depends on the number of branches $f$. One thus concludes, that probability of formation of this types of loops decreases with increasing of number of branches.  For $f=1$, with ${\lambda_1^1}^{{ \rm {Pure}}}$ and ${\lambda_1^2}^{{ \rm {Pure}}} $ one restores the known value of corresponding exponent $\lambda_1$ of a single chain (see Fig.  2). Also, one restores exponent $\lambda_2$ with ${\lambda_3^1}^{{ \rm {Pure}}}$,  ${\lambda_2^2}^{{ \rm {Pure}}}$ and ${\lambda_2^1}^{{ \rm {Pure}}}$ at $f=1$
and exponent $\lambda_3$ with ${\lambda_4^1}^{{ \rm {Pure}}}$ and ${\lambda_3^2}^{{ \rm {Pure}}}$.}

\begin{figure}[t!]
\begin{center}
\includegraphics[width=90mm]{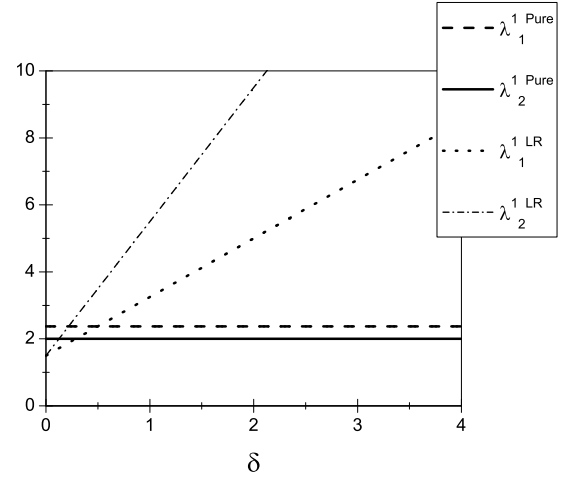}
\caption{ \label{plote} Scaling exponents ${\lambda_1^1}^{{ \rm {Pure}}}$, $ {\lambda_2^1}^{{ \rm {Pure}}}$, ${\lambda_1^1}^{{ \rm {LR}}}$ and ${\lambda_2^1}^{{ \rm {LR}}}$ at fixed $\varepsilon=1$ and $f=5$ as functions of correlation parameter $\delta$. }
\end{center}\end{figure}

Finally, substituting the values of fixed point (\ref{FPL}) into relations (\ref{l1})-(\ref{l6}), we obtain the set of values of critical exponents,
governing the probability of loop formations in polymers in presence of long-range correlated disorder:
\begin{eqnarray}
&&{\lambda_{1}^1}^{\rm  {LR}}=2-\varepsilon/2+\frac{f+2}{4}\delta,\label{exp2}\\
&&{ {\lambda_{2}^1}^{\rm  {LR}}=2-\varepsilon/2+\frac{f+1}{2}\delta,}\label{exp22}\\
&&{\lambda_{3}^1}^{\rm  {LR}}=2-\varepsilon/2+\delta,\\
&&{\lambda_{4}^1}^{\rm  {LR}}=2-\varepsilon/2+\frac{3}{2}\delta,\\
&&{\lambda_{1}^2}^{\rm  {LR}}=2-\varepsilon/2+\frac{3}{4}\delta,\\
&&{\lambda_{2}^2}^{\rm  {LR}}=2-\varepsilon/2+\delta,\\
&&{\lambda_{3}^2}^{\rm  {LR}}=2-\varepsilon/2+\frac{3}{2}\delta.
\end{eqnarray}
{ Again, only exponents ${\lambda_{1}^1}^{\rm  {LR}}$ and ${\lambda_{2}^1}^{\rm  {LR}}$ appear to be $f$-dependent, and increases with  increasing the number of branches.  For $f=1$, with ${\lambda_{1}^1}^{\rm  {LR}}$ and ${\lambda_{1}^2}^{\rm  {LR}} $ one restores the corresponding  $\lambda_1^{\rm  {LR}}$ of a single chain in long-range correlated disorder, found in our previous study \cite{Haydukivska14}. Also, one restores exponent $\lambda_2^{\rm  {LR}}$ with ${\lambda_{3}^1}^{\rm  {LR}}$, ${\lambda_{2}^2}^{\rm  {LR}}$ and ${\lambda_2^1}^{{ \rm {LR}}}$ at $f=1$
and exponent $\lambda_3$ with ${\lambda_{4}^1}^{\rm  {LR}}$ and ${\lambda_{3}^2}^{\rm  {LR}}$.}

Evaluating (\ref{exp1}), (\ref{exp11}), (\ref{exp2})  and (\ref{exp22})  at $d=3$ ($\varepsilon=1$) and various fixed values of $\delta$ and $f$, one finds, that presence of correlated  obstacles leads to an increase of corresponding exponent with increasing the strength of disorder (see Fig. \ref{plote}). Thus, the probability of loop formation of this type in branched star polymers is suppressed in presence of structural disorder.

\section{Conclusions} The loop formation in macromolecules
plays an important role in a number of biochemical processes, such as stabilization of globular proteins or
DNA compactification in the nucleus. In the present study, we addressed the question of how the complex structure of $f$-branched
star polymers influences the statistics of loop formation.
Moreover, since in the real physical situations one often encounters polymers in solutions in presence of complex structural inhomogeneities,
 we consider the star polymers in an environment with
 long-range correlated disorder, where the defects are  correlated at large distances $r$ according to a power law (\ref{corfun}).

Applying the direct polymer renormalization approach, we found analytical expressions for the scaling exponents, governing the probabilities of formation of different types of loops in branched macromolecules (\ref{P}). Note that {  7} different types of loops can be found in star polymers, which are
 formed either by two monomers which belong to the same branch of macromolecule  or to the different branches  (see Fig.  \ref{fig:3}). We found, that
only the probabilities of {  two}  special types of loops in branched polymers, governed by scaling { exponents} ${\lambda_{1}^1}$ {  and ${\lambda_{2}^1}$}, are $f$-dependent and non-trivial as comparing with the case of linear chain.  The presence of correlated  disorder in general leads to an increase of scaling exponents, governing the scaling of probabilities (\ref{P}) and thus, the probability of loop formation of this type in branched star polymers is suppressed in presence of structural defects.

\section*{Appendix A}

\begin{figure}[t!]
\begin{center}
\includegraphics[width=80mm]{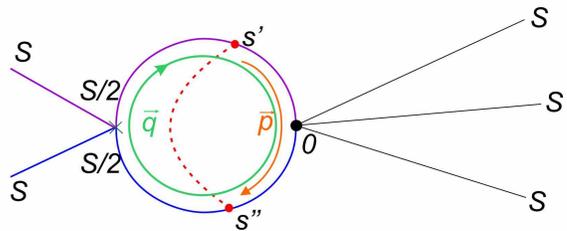}
\caption{ \label{fig:12} Example of diagrammatic contribution into the partition function of star polymer with a loop between two inner monomers of different chains. }
\end{center}\end{figure}

Here, we evaluate an analytical expression corresponding to diagram D65 (see Fig.  \ref{fig:4}) as an example of diagram calculation. More detailed presentation of diagram under consideration is given on figure \ref{fig:12}. Solid lines on a diagram is a schematic presentation of a star polymer with branch length $S$,
dashed line denotes the excluded volume interaction between points $s'$ and $s''$  (interaction points). According to
the general rules of diagram calculations \cite{desCloiseaux}, each segment between any two points $s_1$ and $s_2$
is oriented and bears a wave vector $\vec{q}_{12}$ given by a sum of incoming and outcoming  wave vectors injected
at interaction points and end points. At these points, the flow of wave vectors is
conserved.
A factor $\exp\left(-\frac{\vec{q}_{12}^2}{2}(s_2-s_1)\right)$ is associated with each segment, and integration is to be made
over all independent segment areas and over wave vectors injected at the end points and interaction points.
The diagram shown on Fig.  \ref{fig:12} is then associated with an expression:
\begin{eqnarray}
&&D65=\frac{1}{(2\pi)^{2d}}\int_0^{S/2}\!\!\!\! \!\!\!{\rm d }s'\int_{0}^{S/2}\!\!\!\!\!{\rm d }s'' \int d\vec{q}\! \int d\vec{p}\,\,
{\rm e}^{-\frac{\vec{q}^2}{2}(S-s'-s'')}\nonumber\\
&&\times{\rm e}^{ -\frac{(\vec{q}+\vec{p})^2}{2}(s'+s'')}.\nonumber
\end{eqnarray}
Performing  integration over the wave vectors $\vec{q},\, \vec{p}$ we obtain:
\begin{eqnarray}
&&D65=\frac{1}{(2\pi)^{d}}\int_0^{S/2}\!\!\!\! \!\!\! {\rm d }s'\int_{0}^{S/2}\!\!\!\! \!\!\!{\rm d }s'' (S-s'-s'')^{-d/2}\times\nonumber\\
&&\times(s'+s'')^{-d/2}\nonumber
\end{eqnarray}

Passing to the dimensionless variables $s'=s'/S$,  $s''=s''$, $h=(s'+s'')$, resulting in:
\begin{eqnarray}
&&D65=\frac{S^{2-d}}{(2\pi)^{d}}\int_0^{1/2}\!\! {\rm d }s'\int_{s'}^{1/2+s'}\!\!{\rm d }h (1-h)^{-d/2}(h)^{-d/2}=\nonumber\\
&&=\frac{S^{2-d}}{(2\pi )^{d}}\int_0^{1/2}\!\! {\rm d }s'(B_{1/2+s'}(1-d/2,1-d/2)-\nonumber\\
&&-B_{s'}(1-d/2,1-d/2)),
\end{eqnarray}
where the definition of an incomplete Euler Beta-function $B_s(a,b)=\int_{0}^{s}{\rm d}x\,x^{a-1}(1-x)^{b-1}$ is used.

Making use of relation \cite{Prud}
\begin{eqnarray}
&&\int{\rm d} s s^{\lambda} B_s(a,b)=\frac{s^{\lambda+1}}{\lambda+2}B_s(a,b)-\nonumber\\
&&-\frac{1}{\lambda+1}B_s(a+\lambda+1,b),
\end{eqnarray}
and performing integration over $s'$ we finally receive
\begin{eqnarray}
D65 =  \frac{S^{2-d}}{(2\pi)^{d}}
\left(B\left( 1-d/2 , 2-d/2  \right)+\frac{2^{d-1}}{{d-2}}\right).
\end{eqnarray}

\section*{Appendix B}

Using the rules of diagram calculations given in Appendix A,
we find the analytical expressions corresponding to diagrams shown on Fig.  \ref{fig:4}.
 Note that the prefactor $\frac{1}{(2\pi)^{d}}$ is skipped in expressions below.
 \widetext{
\begin{eqnarray}
&&D11 = S^{2-d} B(1-d/2,2-d/2),\\
&&D12 = S^{2-d}\frac{1}{\left( 1-d/2 \right)\left( 2-d/2\right)},\\
&&D13 = S^{2-d}\frac{{2}^{2-d/2}-2}{\left( 1-d/2 \right)\left( 2-d/2 \right)},\\
&&D14 = S^{2-d}\frac{1}{\left( 1-d/2 \right)} \left( 5 ^{1-d/2} 2 ^{d-2}{\mbox{$_2$F$_1$}\left(1/2,d/2-1;\,3/2;\,1/5\right)}- B(2-d/2,2-d/2)\right),\\
&&D21 = {\left(\frac{S}{2}\right)}^{2-d/2} S^{2-d/2}{\left(\frac{1}{2}\right)}^{2-d/2}B(1-d/2,2-d/2),\\
&&D22 = {\left(\frac{S}{2}\right)}^{-d/2}{S}^{2-d/2}\frac{1}{\left( 1-d/2 \right)} \left( 5 ^{1-d/2} 2 ^{3d/2-4}{\mbox{$_2$F$_1$}\left(1/2,d/2-1;\,3/2;\,1/5\right)}-{\frac {{2}^{3d/2-5}\sqrt {\pi }\Gamma  \left( 2-d/2 \right) }{ \Gamma  \left( 5/2-d/2 \right) }} \right),\\
&&D23 ={\left(\frac{S}{2}\right)}^{2-d}\frac{1}{\left( 1-d/2 \right)\left( 2-d/2 \right)},\\
&&D24 ={\left(\frac{S}{2}\right)}^{-d/2}\frac{{(3S/2)}^{2-d/2}-{S}^{2-d/2}-{(S/2)}^{2-d/2}}{\left( 1-d/2 \right)\left( 2-d/2 \right)},\\
&&D25 ={\left(\frac{S}{2}\right)}^{-d/2}\frac{{S}^{2-d/2}}{\left( 1-d/2 \right)\left( 2-d/2 \right)},\\
&&D26 = {\left(\frac{S}{2}\right)}^{-d/2}\frac{{S}^{2-d/2}\,2 ^{3d/2-4}}{\left( 1-d/2 \right)} \left( 13 ^{1-d/2}{\mbox{$_2$F$_1$}\left(1/2,d/2-1;\,3/2;\,1/13\right)}-\right.\nonumber\\
&&\left.-5 ^{1-d/2}{\mbox{$_2$F$_1$}\left(1/2,d/2-1;\,3/2;\,1/5\right)}\right),\\
&&D27 = {\left(\frac{S}{2}\right)}^{-d/2}\frac{{(2S)}^{2-d/2}-2{S}^{2-d/2}}{\left( 1-d/2 \right)\left( 2-d/2 \right)},\\
&&D31=D35 ={\left(\frac{S}{3}\right)}^{2-d}\frac{1}{\left( 1-d/2 \right)\left( 2-d/2 \right)},\\
&&D32=D34 ={\left(\frac{S}{3}\right)}^{-d/2}\frac{S^{2-d/2}}{\left( 1-d/2\right)}\left(\frac{1}{3}\left(\frac{5}{12}\right)^{1-d/2}\mbox{$_2$F$_1$}\left(1/2,d/2-1;\,3/2;\,1/5\right)-\right.\\
&&\left.-3^{d/2-2}2^{d-3}\sqrt{\pi}
\frac{\Gamma\left( 2-d/2 \right)}{\Gamma\left( 5/2-d/2 \right)}\right),\\
&&D33 ={\left(\frac{S}{3}\right)}^{2-d}B\left( 1-d/2 , 2-d/2 \right),\\
&&D36 ={\left(\frac{S}{3}\right)}^{-d/2}\frac{S^{2-d/2}}{\left( 1-d/2 \right)\left( 2-d/2 \right)},\\
&&D37 ={\left(\frac{S}{2}\right)}^{2-d}\frac{(2S/3)^{2-d/2}-(S/3)^{2-d/2}-S^{2-d/2}}{\left( 1-d/2
 \right)\left( 2-d/2 \right)},\\
&&D38 ={\left(\frac{S}{3}\right)}^{-d/2}\frac{S^{2-d/2}}{\left( 1-d/2 \right)}\left(\frac{7^{1-d/2}}{3\,2^{2-d}}\mbox{$_2$F$_1$}\left(1/2,d/2-1;\,3/2;\,1/21\right)-\right.\nonumber\\
&&\left.-\frac{7^{1-d/2}}{3^{1-d/2}2^{2-d}}\mbox{$_2$F$_1$}\left(1/2,d/2-1;\,3/2;\,1/7\right)\right),\\
&&D39 = {\left(\frac{S}{3}\right)}^{-d/2}\frac{{\left(7S/3\right)}^{2-d/2}-{\left(5S/3\right)}^{2-d/2}-{\left(4S/3\right)}^{2-d/2}+{\left(2S/3\right)}^{2-d/2}}{\left( 1-d/2
 \right)\left( 2-d/2 \right)},\\
&&D310 = {\left(\frac{S}{3}\right)}^{-d/2}\frac{{\left(3S/2\right)}^{2-d/2}-{\left(2S\right)}^{2-d/2}-2\,S^{2-d/2}}{\left( 1-d/2 \right)\left( 2-d/2 \right)}\\
&&D311 = {\left(\frac{S}{3}\right)}^{-d/2}\frac{{2\left(S/3\right)}^{2-d/2}-(2S/3)^{2-d/2}}{\left( 1-d/2
 \right)\left( 2-d/2 \right)},\\
&&D41 =\left( 2\,S \right)^{-d/2} S^{2-d/2}u(2\pi)^{-d/2}\frac{2^{d/2}}{\left( d-2 \right)},\\
&&D42 =\left( 2\,S \right) ^{-d/2} {S}^{2-d/2}\frac{1}{\left( 1-d/2 \right)} \left( \left(\frac{3}{2} \right) ^{1-d/2}{\mbox{$_2$F$_1$}\left(1/2,d/2-1;\,3/2;\,1/3\right)}-{\frac {{2}^{d/2-2}
\sqrt {\pi }\Gamma  \left( 3-d/2 \right) }{ \left( 2-d/2 \right) \Gamma  \left( 5/2-d/2 \right) }} \right),\\
&&D43 = \left( 2\,S \right) ^{-d/2}{S}^{2-d/2}\frac{1}{\left( 1-d/2 \right)\left( 2-d/2 \right)},\\
&&D44 = \left( 2\,S \right) ^{-d/2}{S}^{2-d/2}\frac{{2}^{2-d/2}-2}{\left( 1-d/2 \right)\left( 2-d/2 \right)},\\
&&D45 = \left( 2\,S \right) ^{-d/2}{S}^{2-d/2} \left({\frac {{2}^{d/2}\sqrt {\pi }\Gamma  \left( 1-d/2 \right) }{  \Gamma  \left( 3/2-d/2 \right) }} -{\frac {{2}^{d/2+1}}{ \left( d-2 \right) }} \right),\\
&&D51 = {\left(\frac{3S}{2}\right)}^{-d/2}\frac{{\left(S/2\right)}^{2-d/2}}{\left( 1-d/2 \right)\left( 2-d/2 \right)},\\
&&D52 = {\left(\frac{3S}{2}\right)}^{-d/2}{\left(\frac{S}{2}\right)}^{2-d/2} \left(\frac{{\mbox{$_2$F$_1$}\left(d/2,1-d/2;\,2-d/2;\,1/3\right)}}{\left( 1-d/2 \right)}-\frac{{\mbox{$_2$F$_1$}\left(d/2,2-d/2;\,3-d/2;\,1/3\right)}}{\left( 2-d/2 \right)}\right),\\
&&D53 = {\left(\frac{3S}{2}\right)}^{-d/2}\frac{{S}^{2-d/2}}{\left( 1-d/2 \right)} \left(\left(\frac{7}{8}\right) ^{1-d/2}\left({\frac{3}{4}\mbox{$_2$F$_1$}\left(1/2,d/2-1;\,3/2;\,3/7\right)}-
{\frac{1}{4}\mbox{$_2$F$_1$}\left(1/2,d/2-1;\,3/2;\,1/21\right)}\right)\right.\nonumber\\
&&\left.-\frac{2^{d/2-2}{\mbox{$_2$F$_1$}\left(d/2,2-d/2;\,3-d/2;\,1/3\right)}}{\left( 2-d/2\right)}\right),\\
&&D54 = {\left(\frac{3S}{2}\right)}^{-d/2}\frac{{S}^{2-d/2}}{\left( 1-d/2 \right)} \left(\left(\frac{11}{8}\right) ^{1-d/2}\left({\frac{3}{4}\mbox{$_2$F$_1$}\left(1/2,d/2-1;\,3/2;\,3/11\right)}- \right.\right.\nonumber\\
&&\left.\left.{\frac{1}{4}\mbox{$_2$F$_1$}\left(1/2,d/2-1;\,3/2;\,1/33\right)}\right)-
\frac{2^{d/2-2}{\mbox{$_2$F$_1$}\left(d/2,2-d/2;\,3-d/2;\,1/3\right)}}{\left( 2-d/2\right)}\right),\\
&&D55 = {\left(\frac{3S}{2}\right)}^{-d/2}\frac{{\left(S\right)}^{2-d/2}}{\left( 1-d/2 \right)\left( 2-d/2 \right)},\\
&&D56 = {\left(\frac{3S}{2}\right)}^{-d/2}\frac{{\left(2S\right)}^{2-d/2}-2\,S^{2-d/2}}{\left( 1-d/2 \right)\left( 2-d/2 \right)},\\
&&D57 = {\left(\frac{3S}{2}\right)}^{-d/2}\frac{{\left(11S/6\right)}^{2-d/2}-{\left(7S/6\right)}^{2-d/2}-(5S/6)^{2-d/2}+{\left(S/3\right)}^{2-d/2}}{\left( 1-d/2 \right)\left( 2-d/2 \right)},\\
&&D58 = {\left(\frac{3S}{2}\right)}^{-d/2}{S}^{2-d/2} \left(\frac{{\mbox{$_2$F$_1$}\left(d/2,1-d/2;\,2-d/2;\,2/3\right)}}{\left( 1-d/2 \right)}-
\frac{{\mbox{$_2$F$_1$}\left(d/2,2-d/2;\,3-d/2;\,2/3\right)}}{\left( 2-d/2 \right)}\right),\\
&&D59 = {\left(\frac{3S}{2}\right)}^{-d/2}\frac{{S}^{2-d/2}}{\left( 1-d/2 \right)} \left(\left(\frac{7}{8}\right) ^{1-d/2}\left({\frac{3}{4}\mbox{$_2$F$_1$}\left(1/2,d/2-1;\,3/2;\,3/7\right)}+\right.\right.\nonumber\\
&&+\left.\left.{\frac{1}{4}\mbox{$_2$F$_1$}\left(1/2,d/2-1;\,3/2;\,1/21\right)}\right)-
\frac{{\mbox{$_2$F$_1$}\left(d/2,2-d/2;\,3-d/2;\,1/3\right)}}{\left( 2-d/2\right)}\right),\\
&&D510 = {\left(\frac{3S}{2}\right)}^{-d/2}\frac{{S}^{2-d/2}}{\left( 1-d/2 \right)} \left(\left(\frac{11}{8}\right) ^{1-d/2}\left({\frac{3}{4}\mbox{$_2$F$_1$}\left(1/2,d/2-1;\,3/2;\,3/11\right)} +
\right.\right.\nonumber\\
&&+\left.\left.{\frac{1}{4}\mbox{$_2$F$_1$}\left(1/2,d/2-1;\,3/2;\,1/33\right)}\right)-
\frac{{\mbox{$_2$F$_1$}\left(d/2,2-d/2;\,3-d/2;\,1/3\right)}}{\left( 2-d/2\right)}\right),\\
&&D511 = {\left(\frac{3S}{2}\right)}^{-d/2}{S}^{2-d/2} \left(\frac{\left(3/2\right)^{2-d/2}2^d\sqrt{\pi}}{\left( 1-d\right)\left( 2-d\right)}\frac{\Gamma\left( 2-d/2 \right)}{\Gamma\left( 1/2-d/2 \right)}-
\frac{{\mbox{$_2$F$_1$}\left(d/2,1-d/2;\,2-d/2;\,2/3\right)}}{\left( 1-d/2 \right)}+\right.\nonumber\\
&&\left.\frac{{\mbox{$_2$F$_1$}\left(d/2,2-d/2;\,3-d/2;\,2/3\right)}}{\left( 2-d/2 \right)}-2^{d/2-2}\left(\frac{{\mbox{$_2$F$_1$}\left(d/2,1-d/2;\,2-d/2;\,1/3\right)}}{\left( 1-d/2
 \right)}-\right.\right.\nonumber \\
  && -\left.\left.\frac{{\mbox{$_2$F$_1$}\left(d/2,2-d/2;\,3-d/2;\,1/3\right)}}{\left( 2-d/2 \right)}\right)\right),\\
&&D61 = S^{-d/2}\frac{{\left(S/2\right)}^{2-d/2}}{\left( 1-d/2 \right)\left( 2-d/2 \right)},\\
&&D62 = S^{-d/2}\frac{S^{-d/2}-2{\left(S/2\right)}^{2-d/2}}{\left( 1-d/2 \right)\left( 2-d/2 \right)},\\
&&D63 = S^{2-d}\frac{2^{d-2}}{\left( 2-d \right)},\\
&&D64=D610 = \frac{S^{2-d}}{\left( 1-d/2 \right)}\left(3^{1-d/2}2^{d-3}\mbox{$_2$F$_1$}\left(1/2,d/2-1;\,3/2;\,1/3\right)-2^{d-4}\sqrt{\pi}\frac{\Gamma\left( 3-d/2 \right)}{(2-d/2)\Gamma\left( 5/2-d/2 \right)}\right),\\
&&D65 = S^{2-d}\left(B\left( 1-d/2 , 2-d/2  \right)+\frac{2^{d-1}}{{d-2}}\right),\\
&&D66 = \frac{S^{2-d}}{\left( 1-d/2 \right)}\left(5^{1-d/2}2^{d-3}\mbox{$_2$F$_1$}\left(1/2,d/2-1;\,3/2;\,1/5\right)-2^{4-d}\sqrt{\pi}
\frac{\Gamma\left( 3-d/2 \right)}{\left( 2-d/2 \right)\Gamma\left( 5/2-d/2 \right)}\right),\\
&&D67 = S^{-d/2}\frac{S^{2-d/2}}{\left( 1-d/2 \right)\left( 2-d/2 \right)},\\
&&D68 = S^{-d/2}\frac{{\left(2S\right)}^{2-d/2}-2S^{2-d/2}}{\left( 1-d/2 \right)\left( 2-d/2 \right)},\\
&&D69 = S^{-d/2}\frac{{\left(7S/4\right)}^{2-d/2}-{\left(5S/4\right)}^{2-d/2}-{\left(3S/4\right)}^{2-d/2}+{\left(S/4\right)}^{2-d/2}}{\left( 1-d/2 \right)\left( 2-d/2 \right)}.\\
&&{ D71 = {\left(\frac{S}{2}\right)}^{2-d/2} S^{2-d/2}{\left(\frac{1}{2}\right)}^{2-d/2}B(1-d/2,2-d/2),}\\
&&{ D72 = {\left(\frac{S}{2}\right)}^{-d/2}{S}^{2-d/2}\frac{(1/2)^{2-d/2}}{\left( 1-d/2 \right)} \left( 3 ^{2-d} 2 ^{d-2}{\mbox{$_2$F$_1$}\left(1/2,d/2-1;\,3/2;\,1/9\right)}-{\frac {{2}^{d-3}\sqrt {\pi }\Gamma  \left( 2-d/2 \right) }{ \Gamma  \left( 5/2-d/2 \right) }} \right),}\\
&&{ D73 = {\left(\frac{S}{2}\right)}^{-d/2}\frac{{S}^{2-d/2}\,2 ^{3d/2-4}}{\left( 1-d/2 \right)} \left( 5 ^{1-d/2}{\mbox{$_2$F$_1$}\left(1/2,d/2-1;\,3/2;\,1/5\right)}-\frac {1/2\sqrt {\pi }\Gamma  \left( 2-d/2 \right) }{ \Gamma  \left( 5/2-d/2 \right) }\right)}\\
&&{ D74 ={\left(\frac{S}{2}\right)}^{-d/2}\frac{{S}^{2-d/2}}{\left( 1-d/2 \right)\left( 2-d/2 \right)},}\\
&&{ D75 ={\left(\frac{S}{2}\right)}^{2-d}\frac{1}{\left( 1-d/2 \right)\left( 2-d/2 \right)},}\\
&&{ D76 ={\left(\frac{S}{2}\right)}^{-d/2}\frac{{(3S/2)}^{2-d/2}-{S}^{2-d/2}-{(S/2)}^{2-d/2}}{\left( 1-d/2 \right)\left( 2-d/2 \right)},}\\
&&{ D77 = {\left(\frac{S}{2}\right)}^{-d/2}\frac{{(2S)}^{2-d/2}-2{S}^{2-d/2}}{\left( 1-d/2 \right)\left( 2-d/2 \right)},}\\
\end{eqnarray}
}
Here, $\mbox{$_2$F$_1$}\left(a,b;c;z\right)$ is a hypergeometric function and $\Gamma(a)$ is Euler's Gamma-function.

\end{document}